\documentclass[12pt]{article}
\usepackage[english]{babel}
\usepackage{graphicx,subfigure}
\usepackage{amssymb,amsthm,amsmath,amsfonts}
\usepackage[colorlinks=true,linkcolor=blue,citecolor=red, urlcolor=green]{hyperref} 
\usepackage{cite}

\theoremstyle{plain}

\theoremstyle{definition}

\theoremstyle{remark}
 
 \numberwithin{equation}{section}

\renewcommand{\le}{\leqslant}
\renewcommand{\geq}{\geqslant}

\begin{document}

\title{Two--dimensional equilibrium configurations \\
in Korteweg fluids}

\author{M.~Gorgone${}^{*}$, F.~Oliveri${}^{*}$, A.~Ricciardello${}^{**}$, P.~Rogolino${}^{*}$\\
\ \\
{\footnotesize ${}^{*}$ Department of Mathematical and Computer Sciences,}\\
{\footnotesize Physical Sciences and Earth Sciences, University of Messina}\\
{\footnotesize Viale F. Stagno d'Alcontres 31, 98166 Messina, Italy}\\
{\footnotesize mgorgone@unime.it; foliveri@unime.it, progolino@unime.it}\\
{\footnotesize ${}^{**}$ Faculty of Engineering and Architecture, Kore University of Enna}\\
{\footnotesize Cittadella Universitaria, 94100 Enna, Italy}\\
{\footnotesize angela.ricciardello@unikore.it}
}

\date{Published in \textit{Theoretical and Applied Mechanics}\\ \textbf{49}, 111--122 (2022).}

\maketitle

\begin{abstract}
In this paper, after reviewing the form of the constitutive equations for a third grade Korteweg fluid, recently derived by means of an extended Liu procedure, an equilibrium problem is investigated. 
By considering  a two--dimensional setting, it is derived a single nonlinear elliptic equation such that the equilibrium conditions are identically satisfied. Such an equation is discussed both analytically and numerically. Moreover, by considering a particular boundary value problem of Dirichlet type, some preliminary numerical solutions are presented.
\end{abstract}



\section{Introduction}
\label{sec:intro}

A theory for capillarity, taking into account the interaction phenomena in the presence 
of liquid and vapour phases, has been formulated in a pioneering paper by Korteweg 
\cite{Korteweg}, who in 1901 proposed a constitutive law of Cauchy stress tensor depending 
on first and second order gradients of mass density (see also \cite{Heida-Malek}). The aim 
was to describe the cohesive forces due to long--range interactions among the molecules. In 
particular, in the expression of Cauchy stress tensor introduced by Korteweg it is possible to recognize two contributions, one representing a standard Navier--Stokes term for compressible fluids and one representing a capillarity stress involving the gradients of mass density up to second order. 
In the modern literature, Korteweg--type fluids are referred to as materials of grade 3 (see \cite{Dunn_Raj1995,True_Raja}).

Despite their relevance, this class of fluids  received only a
moderate attention in literature even after the seminal papers by Dunn and Serrin \cite{DunnSerrin,Dunn}, where the compatibility with the basic tenets of rational continuum
thermodynamics \cite{Truesdell} has been extensively studied. In particular, Dunn and Serrin observed that Korteweg fluids are, in general, incompatible with the restrictions 
of the second law of thermodynamics \cite{Truesdell}. To overcome this inconvenient, in \cite{DunnSerrin},  an additional rate of supply of mechanical energy, the interstitial working, suitable to model the long--range interactions between the molecules, has been introduced; in such a way, an energy extra--flux is included in the local 
balance of energy.  A different method for ensuring the compatibility with second law of thermodynamics that has been proposed by M\"uller does not  modify the energy balance with the inclusion of extra--terms, but requires to relax the classical form of the entropy flux by including  an entropy extra--flux \cite{muller}.

In the last years, several authors faced the problem of the compatibility of non-local constitutive laws with entropy principle \cite{CST-JMP-2009,CST-JNET-2010,COP-Elasticity-2011,COP-IJNLM-2013,COP-CMT-2015,OPR-2016,CGOP-2020,GOR-2020,GOR-2021,GR-2022}, and various generalizations have been proposed by introducing  an extension of classical Liu procedure \cite{Liu} for the exploitation of the entropy inequality. 

Recently, in \cite{GorRog2021}, a complete solution set of the thermodynamic
restrictions placed by the entropy principle for third grade Korteweg fluids has been explicitly determined by means of an extended Liu procedure that uses as constraints in the entropy inequality  both the field equations and their gradient extensions up to the order of the derivatives entering the state space. Remarkably, the recovered constitutive functions are compatible with a constraint derived by Serrin \cite{Serrin-1983} guaranteeing that at the equilibrium the phase boundaries are not necessarily restricted to special configurations (spherical, cylindrical, or planar). 

The present paper, moving from the results on Korteweg fluids obtained in \cite{GorRog2021},  aims to investigate, both analytically and numerically, the equilibrium configurations.   
The structure of the paper is as follows. In Section~\ref{sec:model}, we briefly review the  form of the constitutive relations derived in \cite{GorRog2021} for a third grade Korteweg fluid. Then, Section~\ref{sec:equilibrium} concerns with the equilibrium problem in a two--dimensional setting; in particular, it is given a single partial differential equation such that the overdetermined system for the equilibrium of the Korteweg fluid is identically satisfied.
Then, the obtained equilibrium condition is analyzed distinguishing the cases when it reduces to a linear elliptic equation or is fully nonlinear; moreover, considering a boundary value problem of Dirichlet type, some preliminary numerical solutions are presented. Finally,  Section~\ref{sec:conclusions} contains some comments as well as possible future developments.
 
\section{Balance equations}
\label{sec:model}

Let $\mathcal{B}$ be a fluid occupying a compact and simply connected region $\mathcal{C}$ of a Euclidean point space $E^3$; at a continuum level, its evolution  is ruled by the field equations representing the local balances of mass, linear momentum and energy, respectively, 
\begin{equation}
\label{equations}
\begin{aligned}
&\frac{\partial \rho}{\partial t}+\nabla\cdot(\rho\mathbf{v})=0,\\
&\rho\left(\frac{\partial\mathbf{v}}{\partial t}+(\mathbf{v}\cdot\nabla)\mathbf{v}\right)-\nabla\cdot \mathbf{T}=\rho\mathbf{f},\\
&\rho\left(\frac{\partial\varepsilon}{\partial t}+\mathbf{v}\cdot \nabla\varepsilon\right)-\mathbf{T}\cdot\nabla\mathbf{v}+\nabla\cdot\mathbf{q}=\rho\mathbf{f}\cdot\mathbf{v},
\end{aligned}
\end{equation}
where $\rho(t,\mathbf{x})$ is the mass density, $\mathbf{v}(t,\mathbf{x})\equiv(v_1,v_2,v_3)$  the velocity,  $\varepsilon(t,\mathbf{x})$ the internal energy per unit mass, $\mathbf{T}$ the symmetric Cauchy stress tensor, $\mathbf{q}$ the heat flux, and $\mathbf{f}(t,\mathbf{x})$ the external body forces per unit mass; moreover, there are no heat sources.

Field equations~(\ref{equations}) need to be closed by constitutive equations for the Cauchy stress tensor and heat flux in such a way the local entropy production
\begin{equation}
\label{entropy inequality}
\sigma_s=\rho\left(\frac{\partial s}{\partial t}+\mathbf{v}\cdot\nabla s\right)+\nabla\cdot \mathbf{J}
\end{equation}
be non--negative along any admissible thermodynamic process, $s$ being the specific entropy, and $\mathbf{J}$ the entropy flux;  $s$ and $\mathbf{J}$ are constitutive quantities  too. 

A constitutive theory requires the choice of the so called state variables; according to Korteweg, Cauchy stress tensor involves second order gradients of mass density, so that we are in the framework of a second order non--local constitutive theory. 
More precisely,  we analyze the class of Korteweg--type materials 
described by the set of constitutive equations
\begin{equation}
\mathcal{F}=\mathcal{F}^*(\rho, \varepsilon, \nabla \rho, \mathbf{L} , \nabla\varepsilon, \nabla\nabla\rho ),
\end{equation}
where $\mathcal{F}$ is an element of the set $\{\mathbf{T}, \mathbf{q},s,\mathbf{J}\}$, and $\mathbf{L}$ is the symmetric part of velocity gradient.
 
The thermodynamic analysis carried out in  \cite{GorRog2021} moves from the assumptions
\begin{equation}
\label{constitutiveassumptions}
\begin{aligned}
\mathbf{T}&=\left(-p+\alpha_1\Delta\rho+\alpha_2|\nabla\rho|^2\right)\mathbf{I}+\alpha_3\nabla\rho\otimes\nabla\rho\\
&+\alpha_4\nabla\nabla\rho+\alpha_5(\nabla\cdot \mathbf{v})\mathbf{I}
+\alpha_6 \mathbf{L},\\
\mathbf{q}&=q^{(1)}\nabla\varepsilon+q^{(2)}\nabla\rho,
\end{aligned}
\end{equation}
where $p$, $\alpha_i$ ($i=1,\ldots,6$) and $q^{(i)}$ ($i=1,2$) are suitable material functions depending on the mass density $\rho$ and the internal energy $\varepsilon$; moreover, the specific entropy $s$ is expanded around the homogeneous state (where all gradients vanish) retaining only the lower order terms in the gradients of mass density and internal energy.

The compatibility with the second principle of thermodynamics, through the use of an extended Liu procedure \cite{cim07}, allows the authors to obtain:
\begin{itemize}
\item  $s=s_0(\rho,\varepsilon)+s_1(\rho)|\nabla\rho|^2$, 
where $s_0$ represents the equilibrium entropy defined for homogeneous states; moreover, in order to satisfy the principle of maximum entropy at the equilibrium, it has to be $s_1(\rho)\le 0$;
\item $\displaystyle \mathbf{J}=\mathbf{q}\frac{\partial s_0(\rho,\varepsilon)}{\partial \varepsilon}+2\rho^2 s_1(\rho)(\nabla\cdot\mathbf{v})\nabla\rho$;
\item the following expressions for the material functions entering the Cauchy stress tensor:
\begin{equation}
\label{constitutive_T}
\begin{aligned}
&p(\rho,\varepsilon)=-\rho^2\frac{\partial s_0}{\partial \rho}\left(\frac{\partial s_0}{\partial \varepsilon}\right)^{-1},\qquad
\alpha_1(\rho,\varepsilon)=-2\rho^2 s_1\left(\frac{\partial s_0}{\partial \varepsilon}\right)^{-1},\\
&\alpha_2(\rho,\varepsilon)=-\rho\left(\frac{\partial s_0}{\partial \varepsilon}\right)^{-1}\left(\rho\frac{\partial s_1}{\partial \rho}+2s_1\right),\\
&\alpha_3(\rho,\varepsilon)=2\rho s_1\left(\frac{\partial s_0}{\partial \varepsilon}\right)^{-1},\qquad
\alpha_4=0.
\end{aligned}
\end{equation}
\end{itemize}

Furthermore, the physically admissible constraints 
\begin{equation}
\begin{aligned}
&q^{(1)}\frac{\partial^2 s_0}{\partial \varepsilon^2}\geq0,\qquad q^{(2)}\frac{\partial^2 s_0}{\partial \rho\partial \varepsilon}\geq0,\qquad \alpha_5\frac{\partial s_0}{\partial \varepsilon}\geq0,\qquad \alpha_6\frac{\partial s_0}{\partial \varepsilon}\geq0,
\end{aligned}
\end{equation}
together with
\begin{equation}
\label{relationq1q2}
q^{(1)}\frac{\partial^2 s_0}{\partial\rho \partial\varepsilon}-q^{(2)}\frac{\partial^2 s_0}{\partial \varepsilon^2}=0,
\end{equation}
need to be satisfied.

Defining at thermodynamical equilibrium the absolute temperature $\theta$ by the classical relation 
$\displaystyle\frac{1}{\theta}=\frac{\partial s_0}{\partial\varepsilon}$,
under the hypothesis  $\displaystyle\frac{\partial^2 s_0}{\partial\varepsilon^2}\neq 0$, the internal energy $\varepsilon$ can be thought of as a function of $\rho$ and $\theta$, \emph{i.e.}, 
$\varepsilon=\varepsilon(\rho,\theta)$.

Finally, using (\ref{relationq1q2}), it is 
\begin{equation}
\label{relationq1q2bis}
q^{(2)}=-q^{(1)}\frac{\partial\varepsilon}{\partial\rho},
\end{equation}
and the heat flux turns out to be expressed by the classical Fourier law
\[
\mathbf{q} = q^{(1)}\frac{\partial\varepsilon}{\partial\theta}\nabla\theta.
\]
The above results allow to rewrite the entropy flux $\mathbf{J}$, and recognize the  classical term $\displaystyle \frac{\mathbf{q}}{\theta}$ and an entropy extra-flux \cite{muller}. 

In the next Section, we consider the equations for a Korteweg fluid in two space dimensions; more in detail, assuming the fluid to be in a vertical plane and subject to gravity, we investigate the equilibrium configurations.

\section{Equilibrium problem}
\label{sec:equilibrium}

By using the solution to the constitutive functions provided in the previous Section, let us study the equilibrium problem on a purely mechanical framework. 

The search for equilibrium configurations of a Korteweg--type fluid consists in finding solutions of the following condition:
\begin{equation} 
\label{korteq}
\nabla\cdot\left(\left(-p+\alpha_1\Delta\rho+\alpha_2|\nabla\rho|^2\right) \mathbf{I}
+\alpha_3\nabla\rho\otimes\nabla\rho\right)+\rho\mathbf{g}=\mathbf{0},
\end{equation}
where $\mathbf{g}$ is the gravity acceleration, whereas $p$, $\alpha_i$ ($i=1,\dots,3$), given in (\ref{constitutive_T}), depend only on $\rho$ and need to be evaluated at constant temperature. 

Let the Korteweg fluid be in the plane $xy$ with $y$ axis directed along the ascending vertical. The equilibrium condition (\ref{korteq}) reads:
\begin{equation}
\label{equilibriumcond}
\begin{aligned}
&\left(-p+\alpha_1(\rho_{xx}+\rho_{yy})+\alpha_2(\rho_x^2+\rho_y^2)+\alpha_3\rho_x^2\right)_x+\left(\alpha_3\rho_x\rho_y\right)_y=0,\\
&\left(\alpha_3\rho_x\rho_y\right)_x+\left(-p+\alpha_1(\rho_{xx}+\rho_{yy})+\alpha_2(\rho_x^2+\rho_y^2)+\alpha_3\rho_y^2\right)_y-\rho g=0,
\end{aligned}
\end{equation}
where the subscripts ${(\cdot)}_x$ and ${(\cdot)}_y$ stand for partial derivatives with respect to the indicated variables, and $g$ is the modulus of gravity acceleration. We observe that equilibrium conditions (\ref{equilibriumcond}) represent an overdetermined system of two partial differential equations in the unknown $\rho(x,y)$.

A theorem by Serrin \cite{Serrin-1983}, based on a result by Pucci \cite{Pucci}, states that, unless rather special conditions on the coefficients involved in (\ref{equilibriumcond}) are satisfied, only very simple geometric phase boundaries (spherical, cylindrical, or
planar) are admitted.

In fact,  in order to have the possibility to have more general geometric phase boundaries at equilibrium, it is necessary that the constitutive quantities involved in the Cauchy stress tensor satisfy the following condition:
\begin{equation}
\label{serricond}
\alpha_3^2-\alpha_1\frac{\partial\alpha_3}{\partial\rho}+2\alpha_2\alpha_3=0.
\end{equation}

It is worth of observing that  condition (\ref{serricond}) is not physically necessary, in the sense that, although rather unusual, without admitting it  very few equilibrium configurations are allowed; remarkably, the constitutive relations deduced in \cite{GorRog2021} satisfy this condition, provided that
\begin{equation}\label{serrin_result}
s_0(\rho,\varepsilon)=s_{01}(\rho)+s_{02}(\varepsilon),
\end{equation}
where $s_{01}$ and $s_{02}$ are functions of the indicated arguments.

Furthermore, condition (\ref{serrin_result}), from relations (\ref{relationq1q2}) and (\ref{relationq1q2bis}), implies  $\varepsilon=\varepsilon(\theta)$, \emph{i.e.}, the internal energy depends only upon the absolute temperature, and the heat flux becomes
\begin{equation}
\mathbf{q} = q^{(1)}\frac{d\varepsilon}{d\theta}\nabla\theta.
\end{equation}

In the light of previous considerations, after simple algebraic manipulations, the condition
\begin{equation}
\label{risolvente}
2\rho s_1 \left(\rho_{xx}+\rho_{yy}\right) +\frac{d(\rho s_1)}{d\rho}\left(\rho_x^2+\rho_y^2\right)-\frac{d(\rho s_{01})}{d\rho}+\frac{g}{\theta_0}y-\kappa=0,
\end{equation}
where $s_{01}$ and $s_1$ are functions of $\rho$, $\theta_0$ is the constant absolute temperature at the equilibrium, and $\kappa$ is an arbitrary integration constant, can be obtained; it represents  the only equation to be solved in order to identically satisfy conditions (\ref{equilibriumcond}) and so find the equilibrium configurations.

\subsection{Equilibrium configurations}
Hereafter, we present some preliminary results, both from analytical and numerical viewpoints, about the equilibrium configurations, and exhibit some solutions.

At first, we have to choose the functional expression of the constitutive quantities $s_{01}(\rho)$ and $s_1(\rho)$. Let us assume
\begin{equation}
s_{01}=\kappa_1 \rho^m,\qquad s_1=-\kappa_2 \rho^n, 
\end{equation}
with $\kappa_i\in\mathbb{R}^{+}$ ($i=1,2$), and $m, n\in\mathbb{R}$; then, equation (\ref{risolvente}) becomes
\begin{equation}
\label{risolvente-bis}
2\kappa_2\rho^{n+1}\left(\rho_{xx}+\rho_{yy}\right) +\kappa_2(n+1)\rho^n\left(\rho_x^2+\rho_y^2\right)+\kappa_1(m+1)\rho^m-\frac{g}{\theta_0}y+\kappa=0.
\end{equation}

Let us fix in the plane $xy$ the rectangular domain
$[0,\ell_1]\times [0,\ell_2]$ ($\ell_1,\ell_2> 0$)
where the equation will be studied.

Introducing dimensionless variables by the substitutions
\[
x\rightarrow \ell_1 x,\qquad y\rightarrow \ell_1 y,\qquad \rho\rightarrow  R_0 \rho,
\]
$R_0$ being a reference density,  equation (\ref{risolvente-bis}) writes
\begin{equation}
\label{dimensionless}
\rho^{n+1}\left(\rho_{xx}+\rho_{yy}\right) +\frac{n+1}{2}\rho^n\left(\rho_x^2+\rho_y^2\right)+\alpha(m+1) \rho^m+\beta y+\gamma=0,
\end{equation}
where
\[
\alpha=\frac{\kappa_1}{2\kappa_2}\ell_1^2R_0^{m-n-2},\quad
\beta=-\frac{g }{2\kappa_2\theta_0}\ell_1^3R_0^{-n-2},\quad
\gamma=\frac{\kappa}{2\kappa_2}\ell_1^2R_0^{-n-2},
\]
that we study in the domain 
\[
\Omega=[0,1]\times [0,d],\qquad d=\frac{\ell_2}{\ell_1},
\]
with Dirichlet boundary conditions
\[
\rho(x,0)=g_1(x), \quad \rho(0,y)=g_2(y),\quad \rho(x,d)=g_3(x),\quad \rho(1,y)=g_4(y),
\]
where the smooth functions $g_1(x)$, $g_2(y)$, $g_3(x)$ and $g_4(y)$ will be specified below.

Equation (\ref{dimensionless}) is a nonlinear elliptic partial differential equation that, in the special cases where  $m=\pm 1$ and $n=-1$, becomes linear.
In fact, when $m=1$ and $n=-1$, the condition for equilibrium (\ref{dimensionless}) reads
\begin{equation}
\label{m=1}
\rho_{xx}+\rho_{yy}+2\alpha\rho+\beta y+\gamma=0,
\end{equation}
that, using the transformation
\[
\rho=\overline{\rho}-\frac{\beta y+\gamma}{2\alpha},
\]
becomes
\[
\overline{\rho}_{xx}+\overline{\rho}_{yy}+2\alpha\overline{\rho}=0,
\]
that is a Poisson equation for which many analytical solutions can be found, for instance in separable form. 

On the contrary, if $m=-1$ and $n=-1$, equation (\ref{dimensionless}) becomes
\[
\rho_{xx}+\rho_{yy}+\beta y+\gamma=0,
\]
that, through the transformation
\[
\rho=\overline{\rho}-\frac{\beta}{6}y^3-\frac{\gamma}{2}y^2,
\]
reduces to the Laplace equation
\[
\overline{\rho}_{xx}+\overline{\rho}_{yy}=0.
\]

\begin{figure}
\centering
\includegraphics[width=0.43\textwidth]{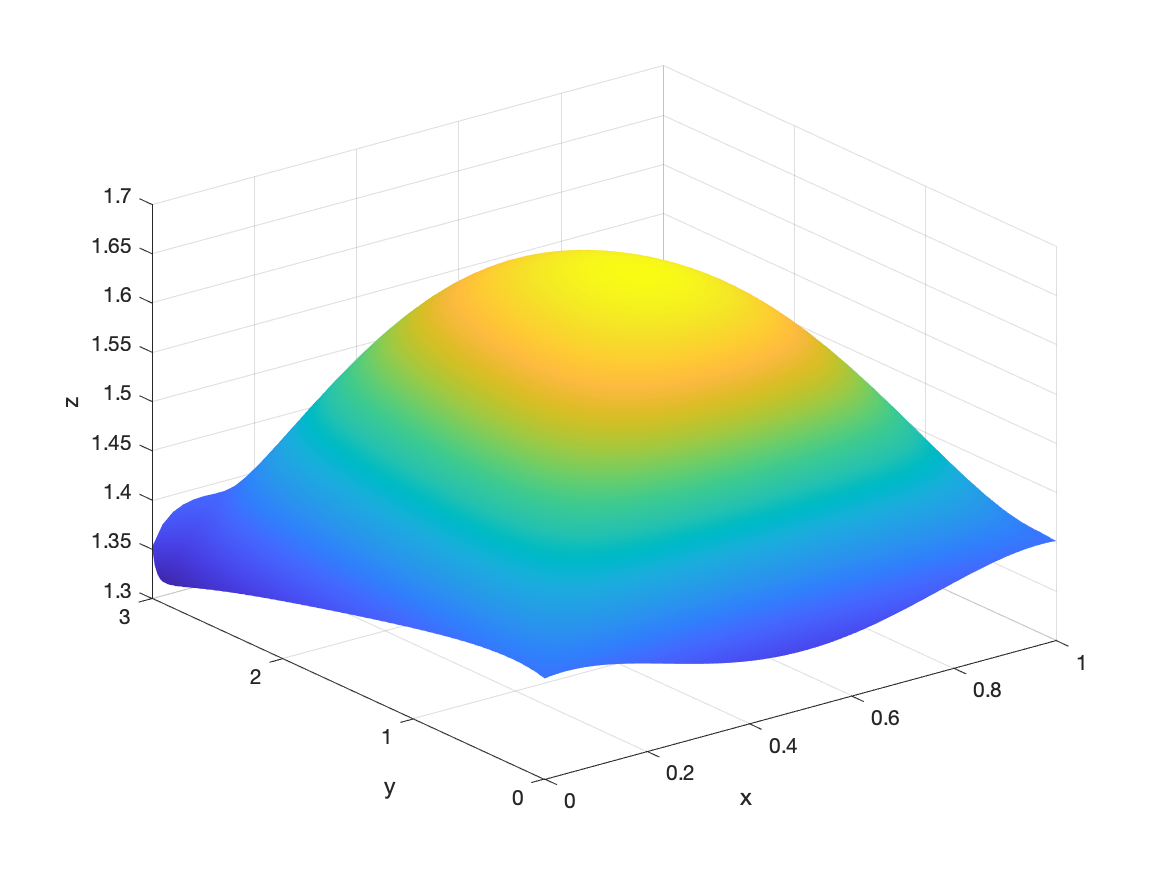}
\includegraphics[width=0.43\textwidth]{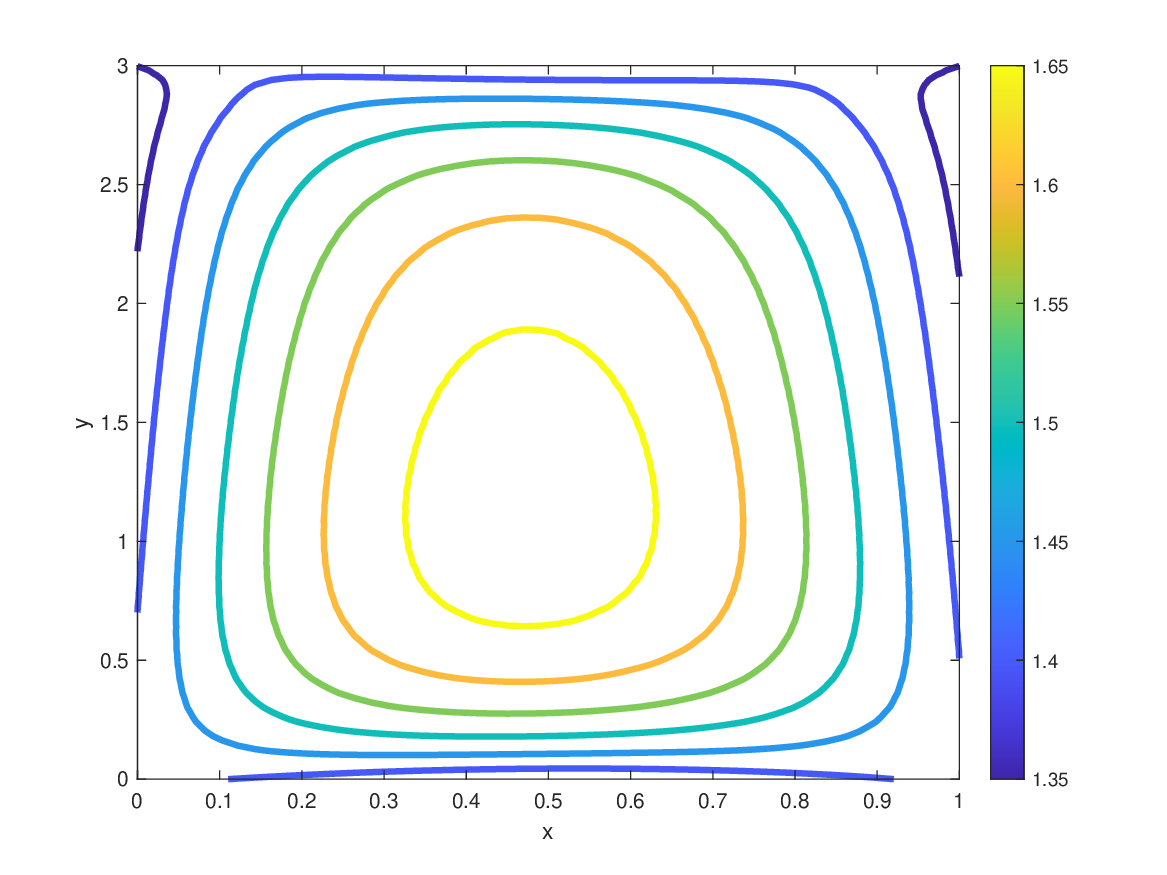}
\includegraphics[width=0.43\textwidth]{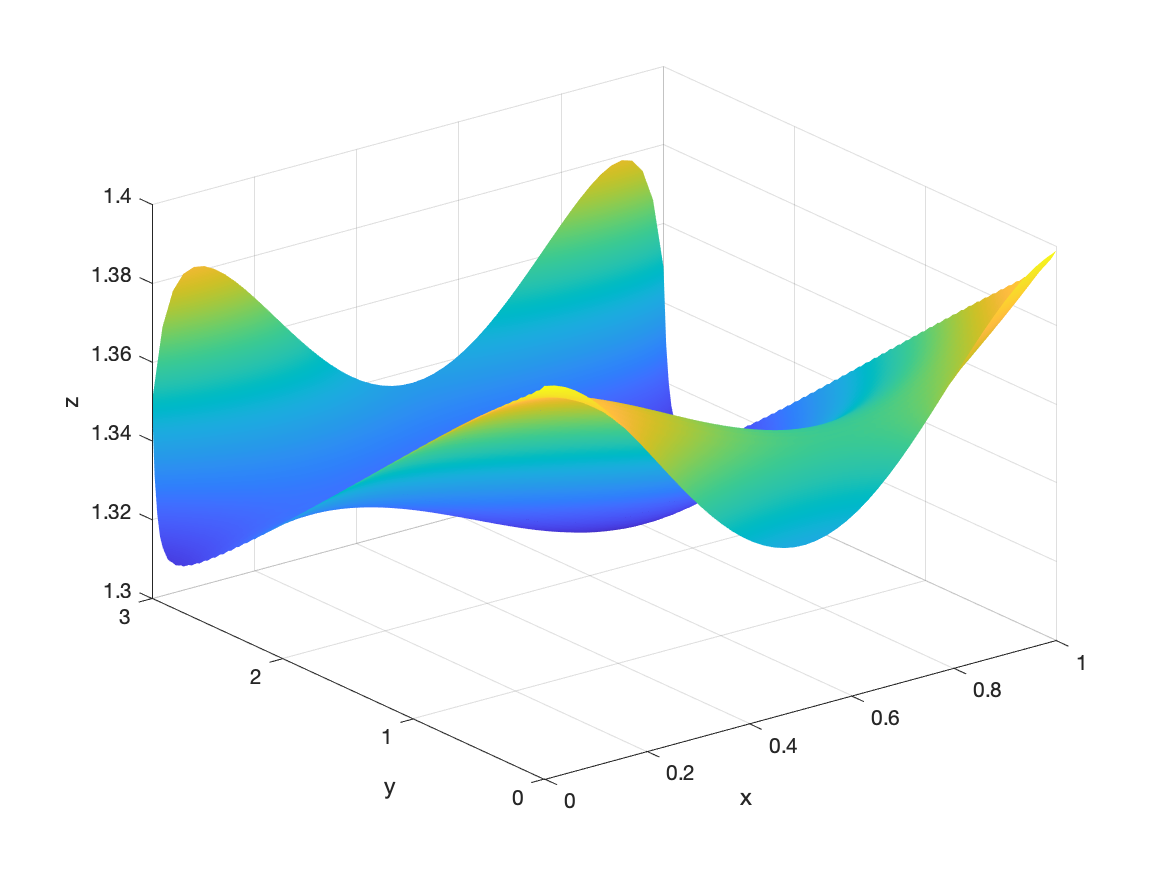}
\includegraphics[width=0.43\textwidth]{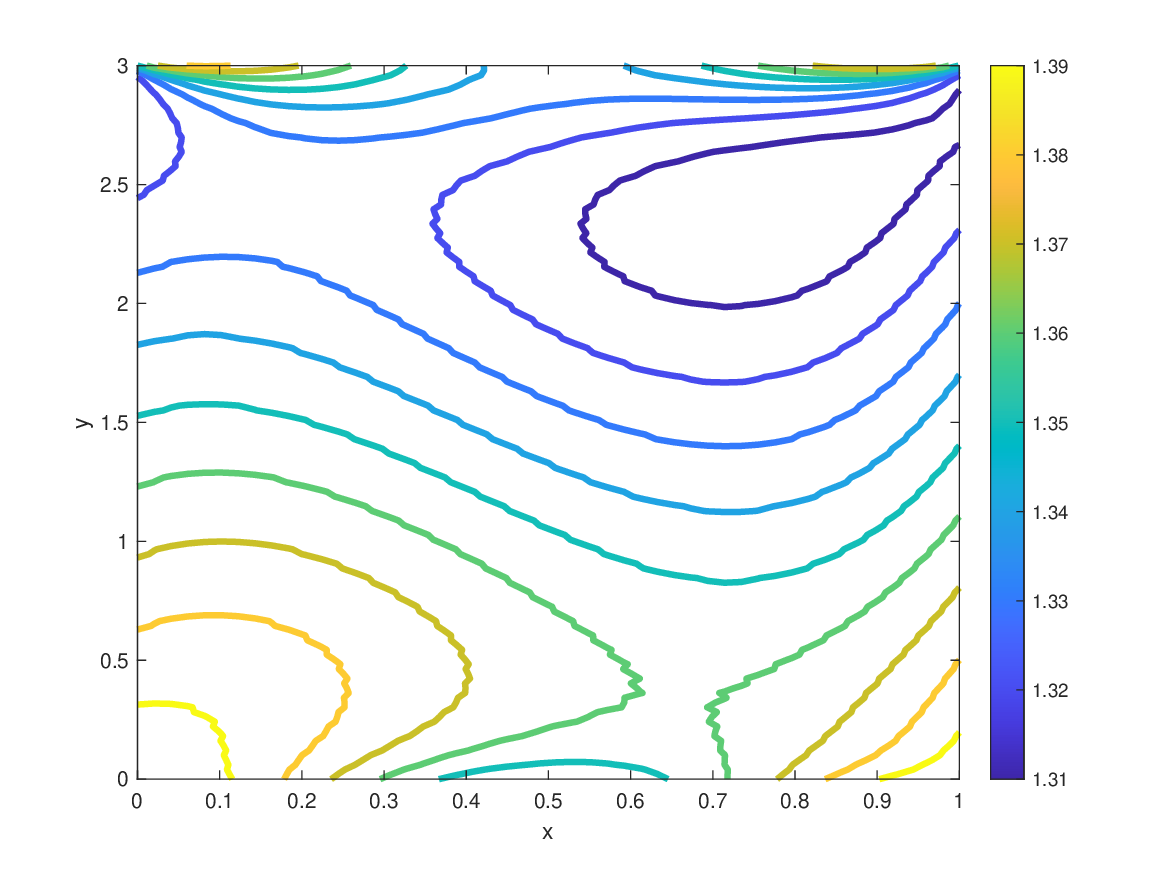}
\includegraphics[width=0.43\textwidth]{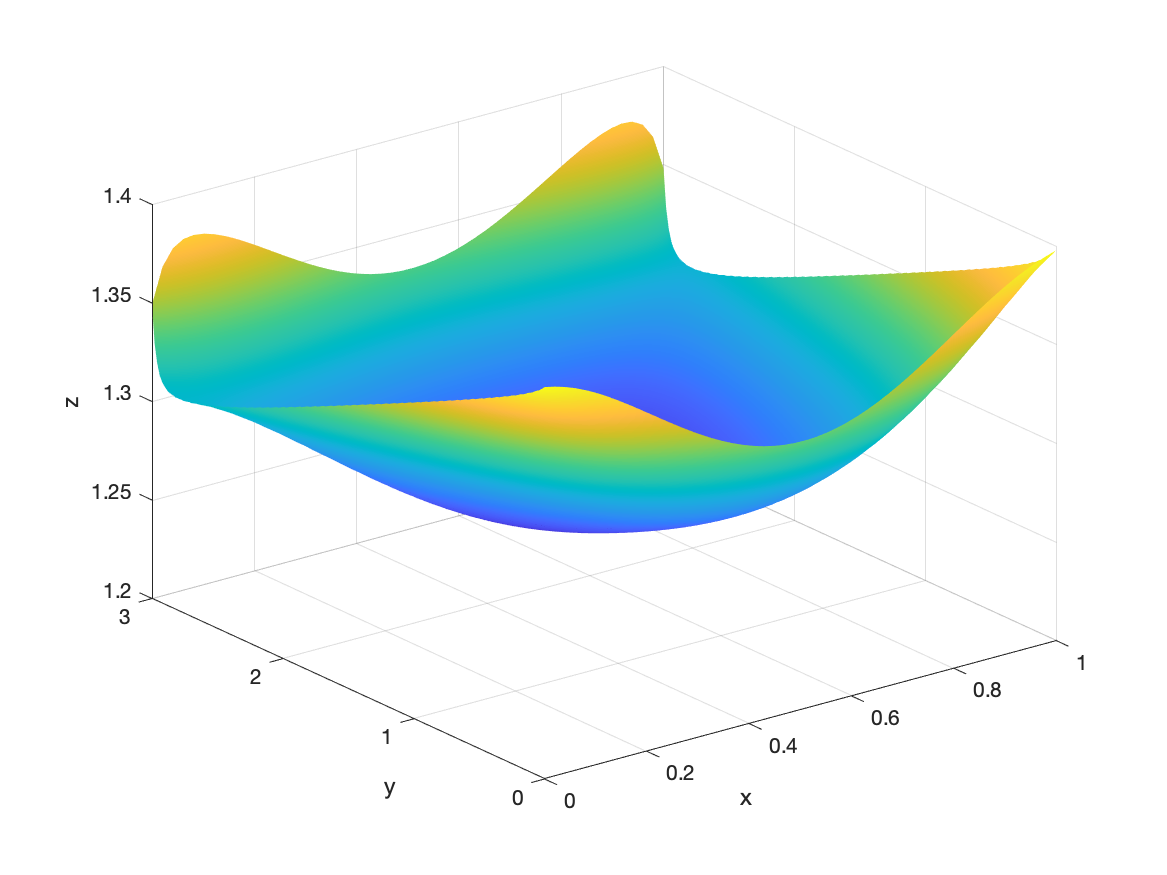}
\includegraphics[width=0.43\textwidth]{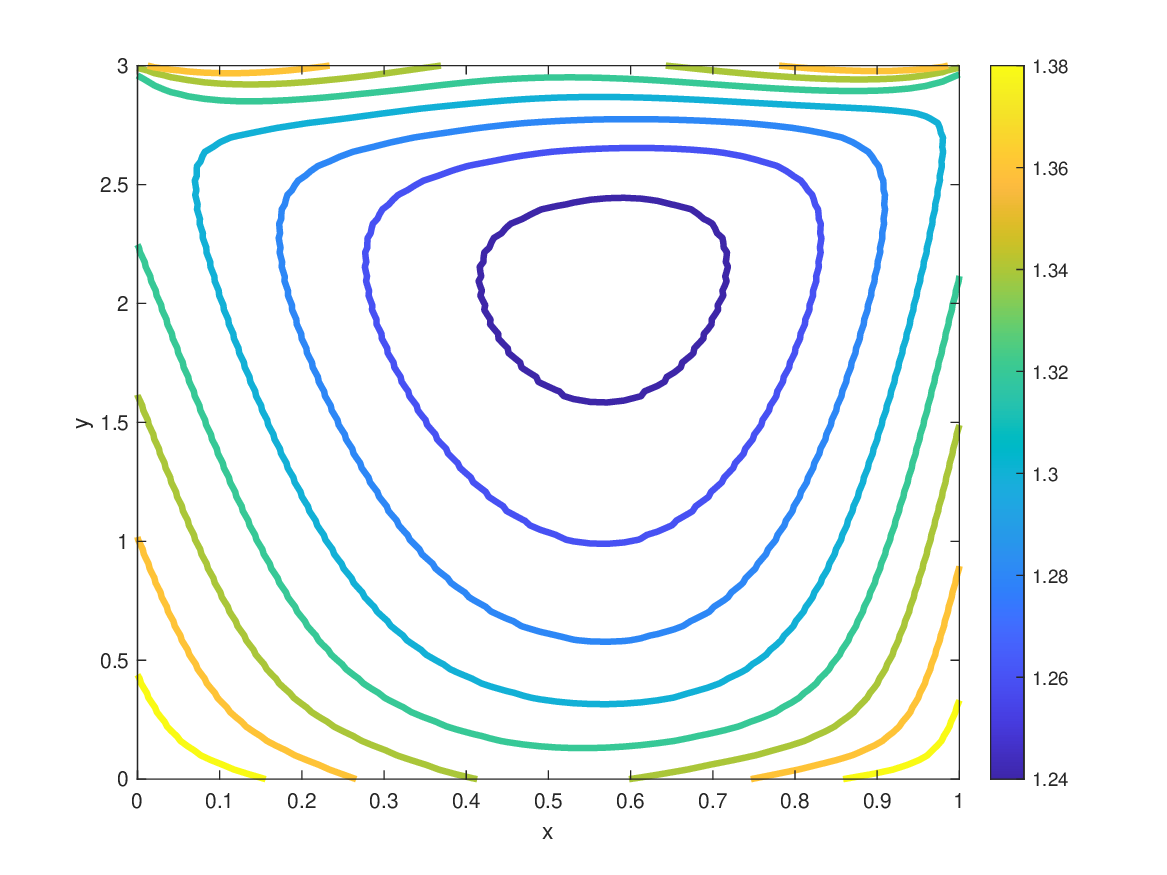}
\includegraphics[width=0.43\textwidth]{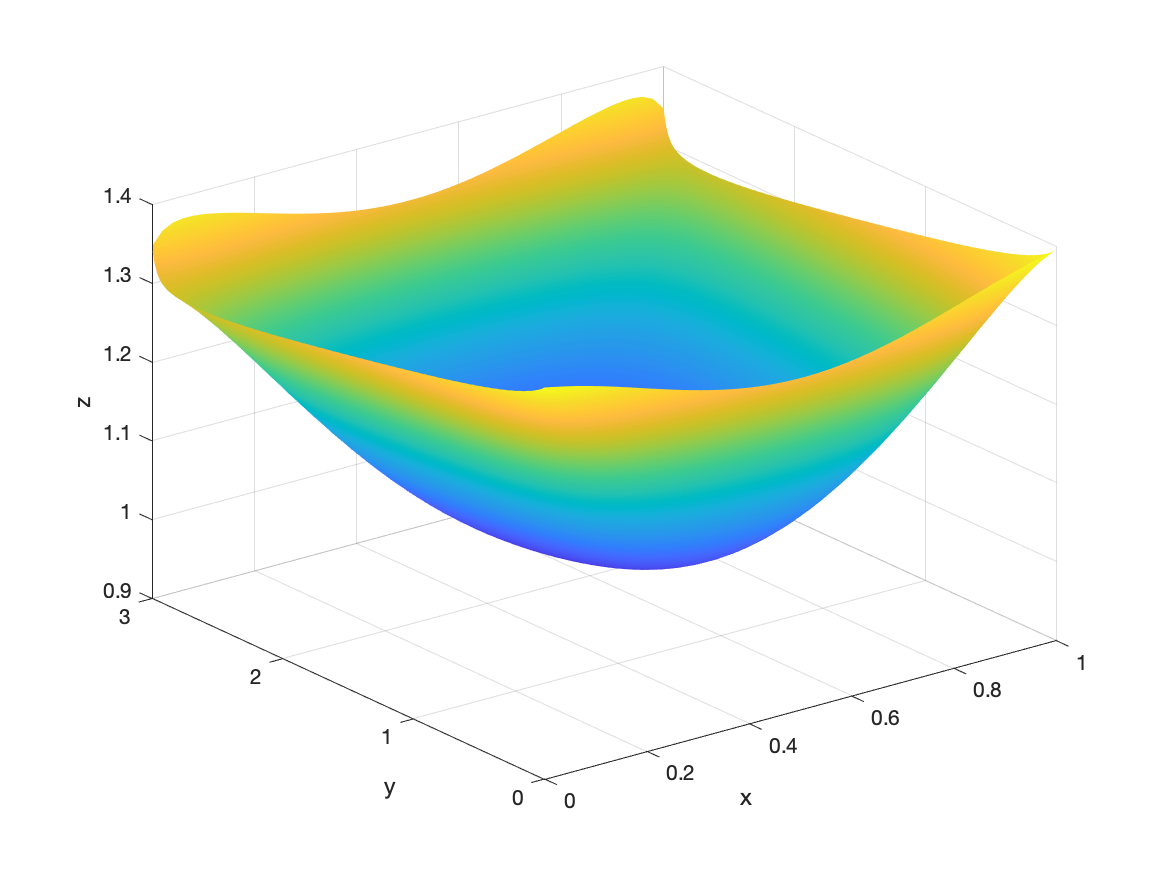}
\includegraphics[width=0.43\textwidth]{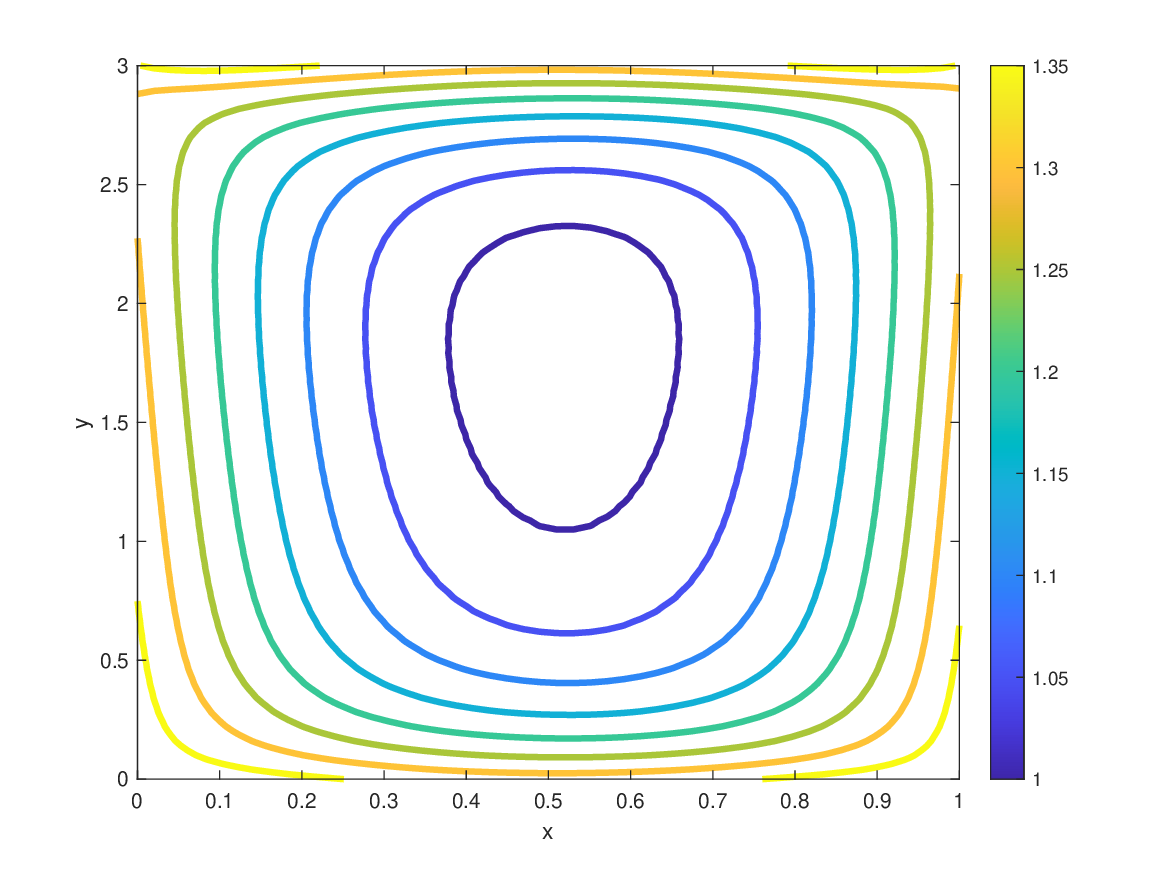}
\caption{Plot of the density $\rho$ (left) and contour plot (right). The values of the parameters are (from the top): ($m=1, \gamma=1$), ($m=1, \gamma=-1$), ($m=-1, \gamma=1$), ($m=-1, \gamma=-1$).}
\label{fig1}
\end{figure}

\begin{figure}
\centering
\includegraphics[width=0.43\textwidth]{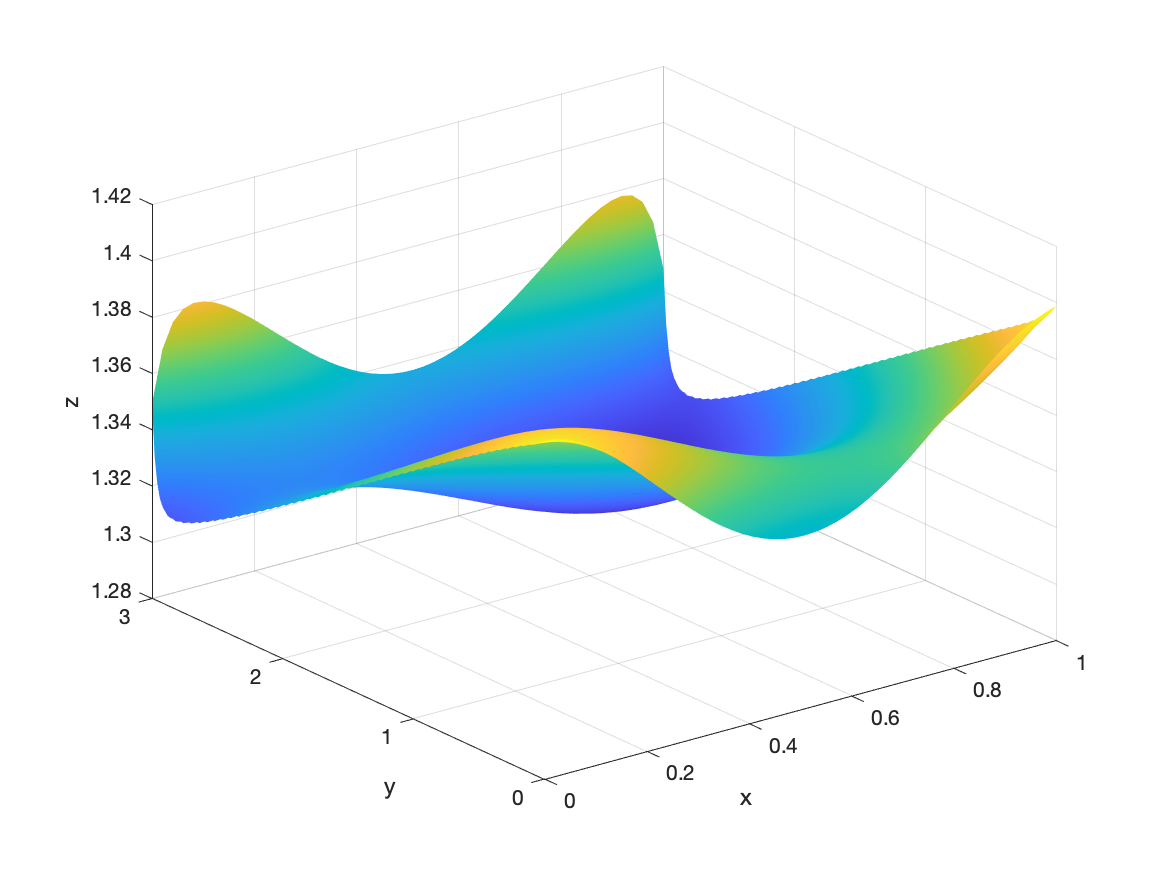}
\includegraphics[width=0.43\textwidth]{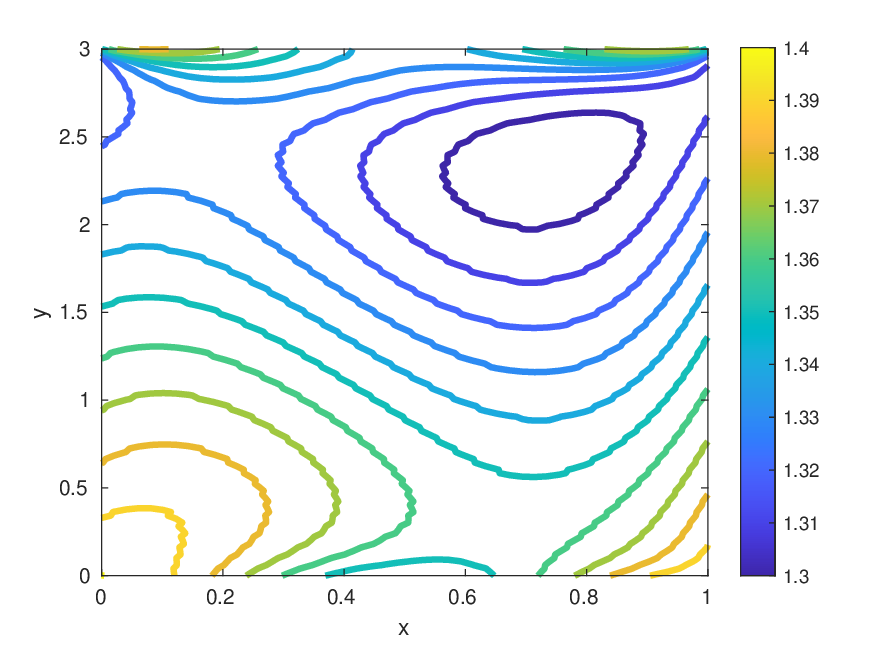}
\includegraphics[width=0.43\textwidth]{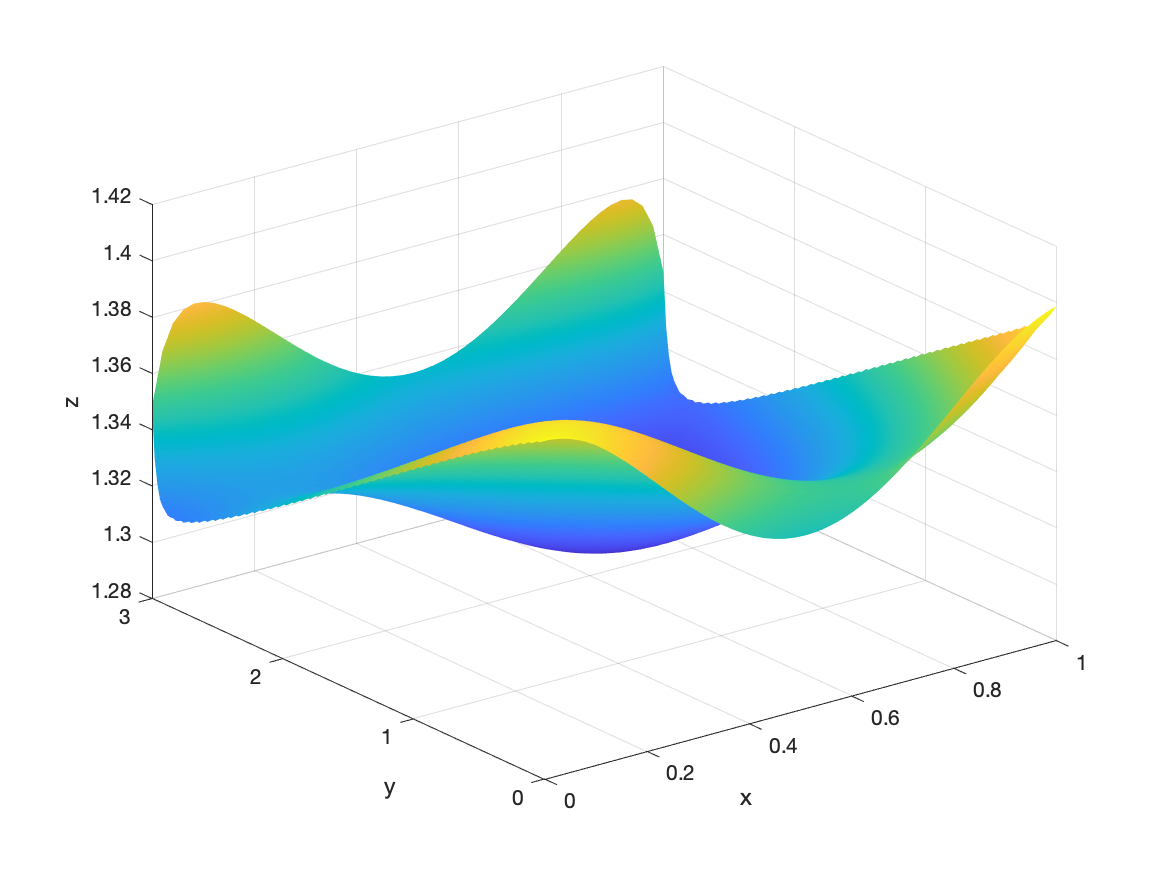}
\includegraphics[width=0.43\textwidth]{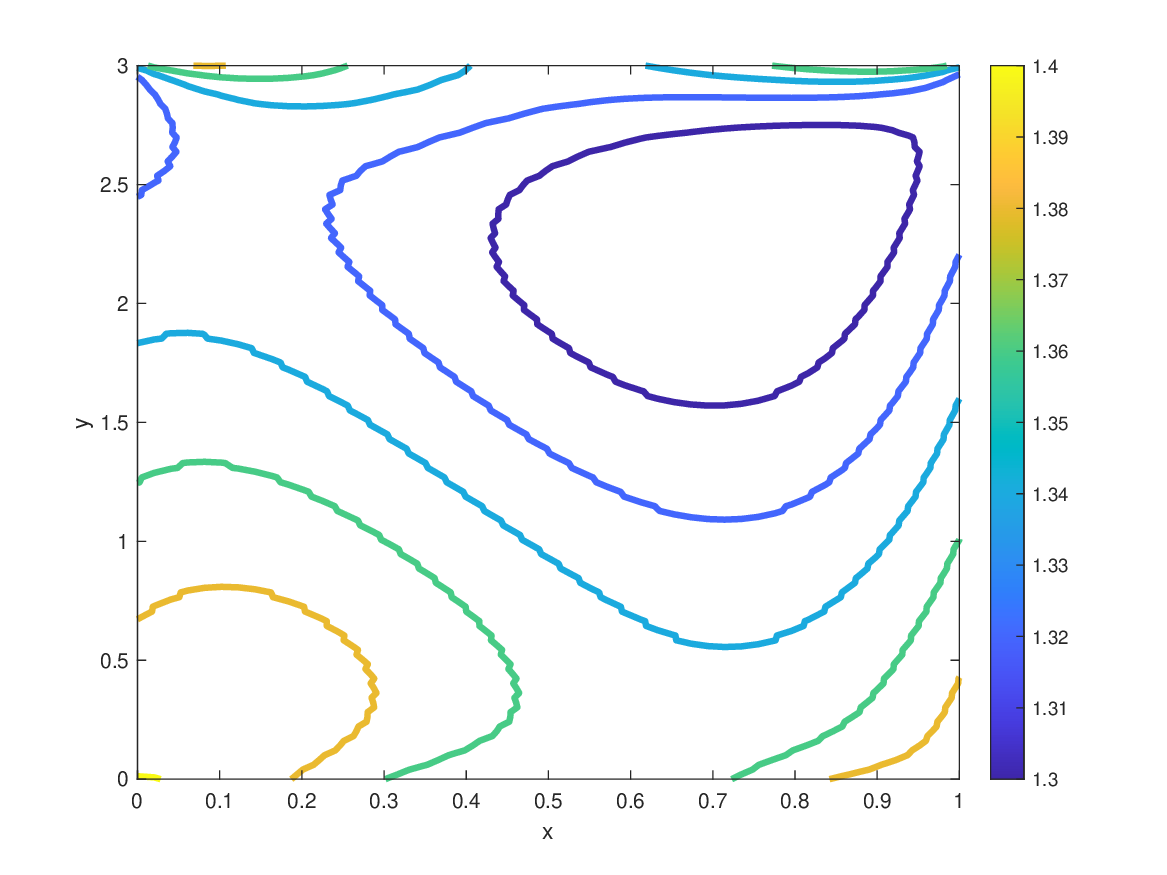}
\includegraphics[width=0.43\textwidth]{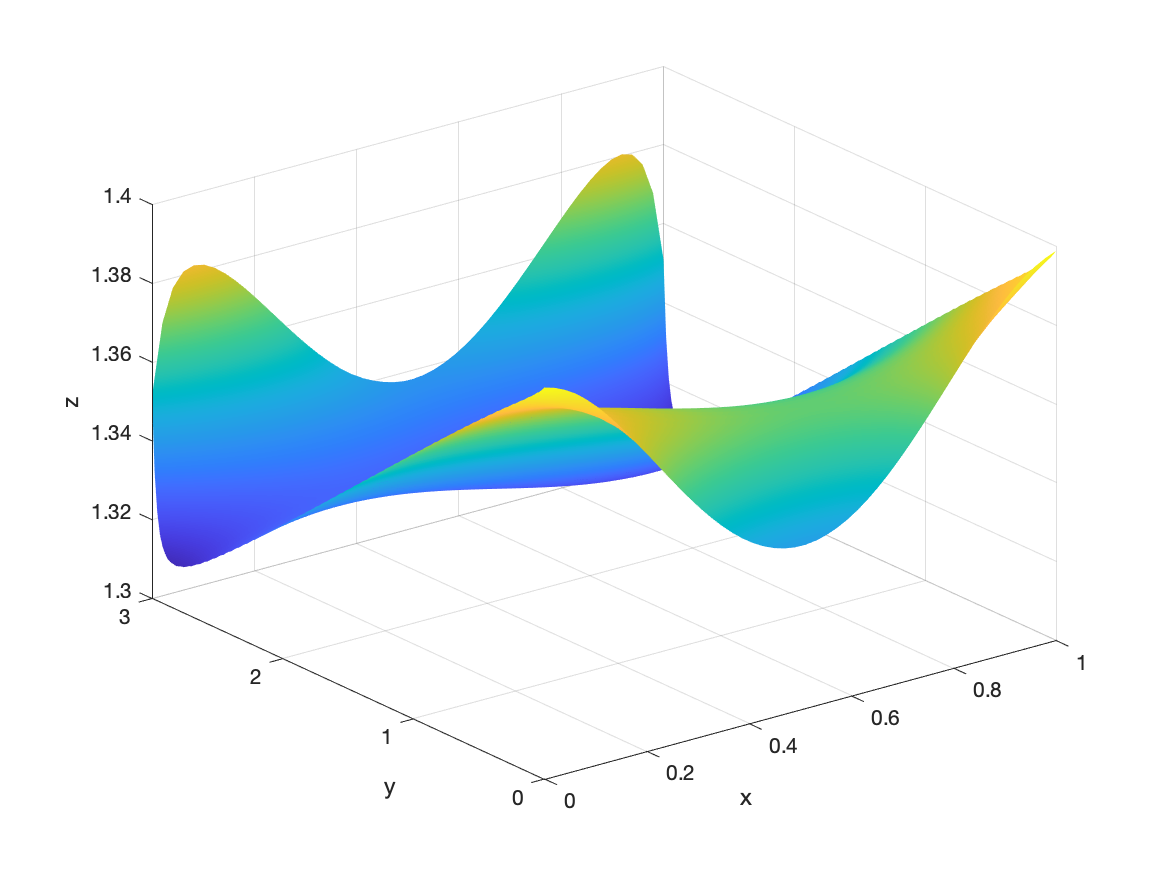}
\includegraphics[width=0.43\textwidth]{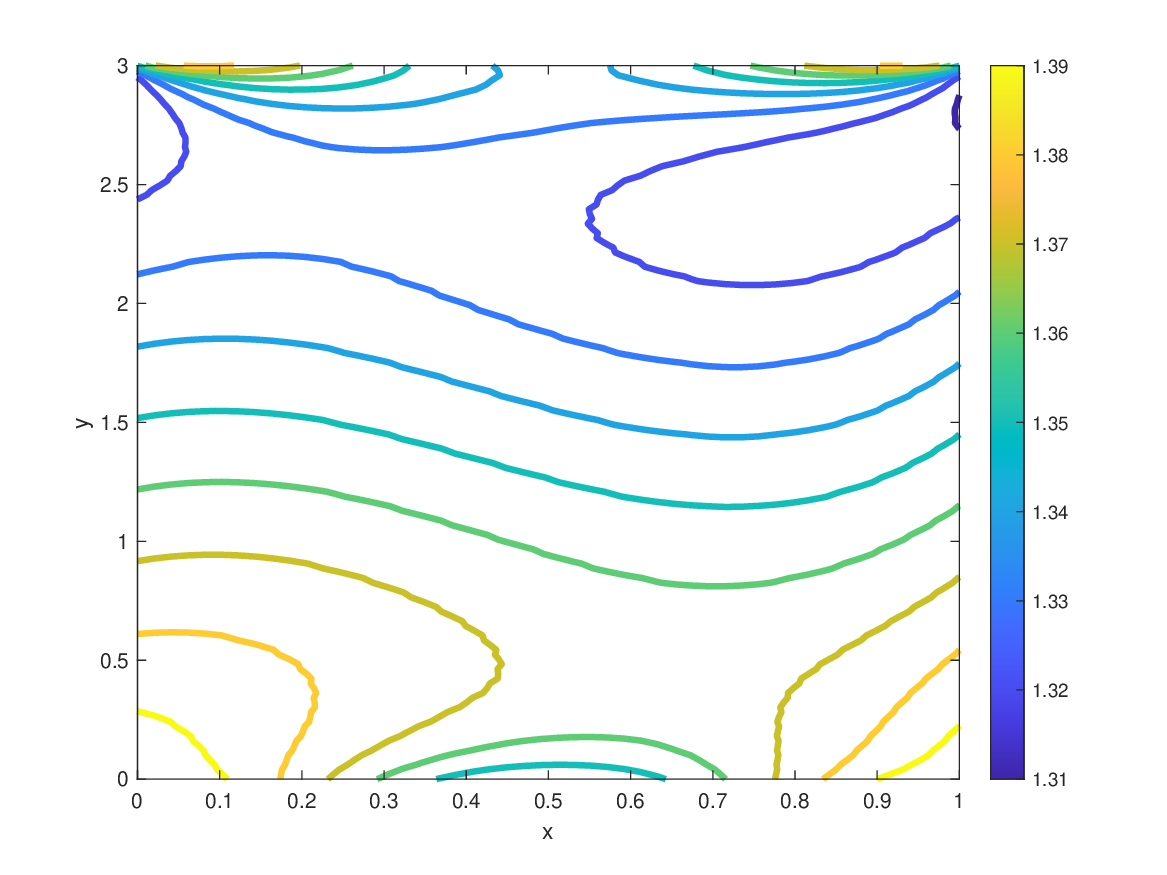}
\includegraphics[width=0.43\textwidth]{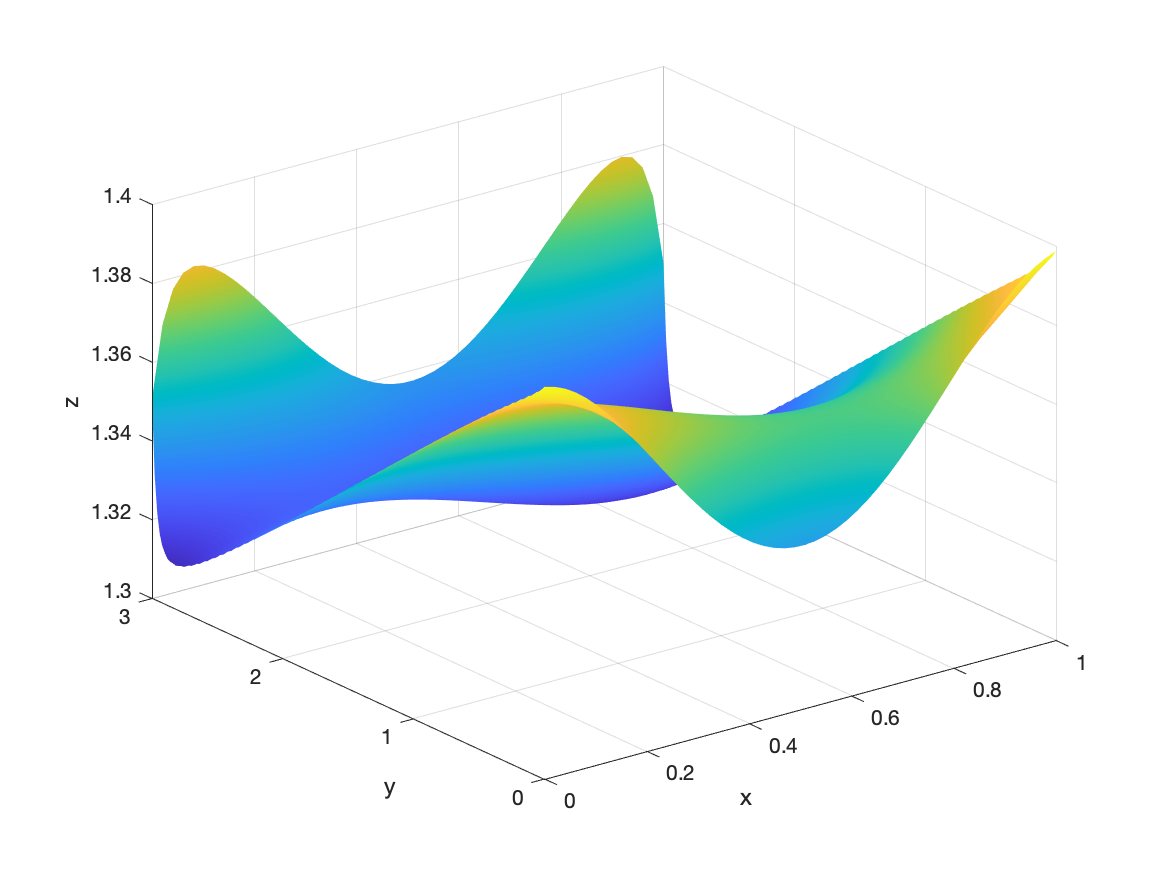}
\includegraphics[width=0.43\textwidth]{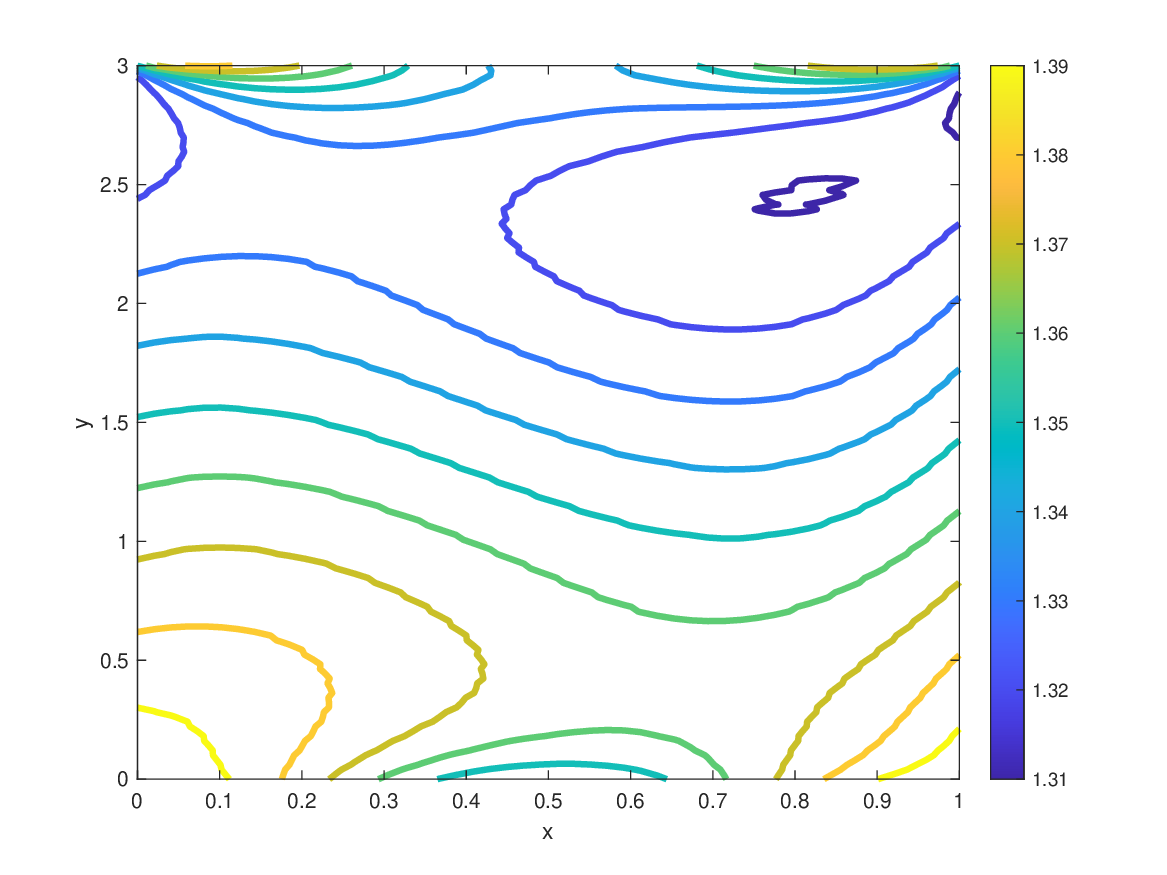}
\caption{Plot of the density $\rho$ (left) and contour plot (right). The values of the parameters are (from the top): ($m=1, n=-2$), ($m=1, n=-3$), ($m=1, n=1$), ($m=1,n=0$).}
\label{fig2}
\end{figure}

Let us now consider the following boundary value problem:
\begin{equation}
\label{BVP}
\left\{
\begin{aligned}
&\rho^{n+1}\left(\rho_{xx}+\rho_{yy}\right) +\frac{n+1}{2}\rho^n\left(\rho_x^2+\rho_y^2\right)+\alpha(m+1) \rho^m+\beta y+\gamma=0,\\
&(x,y)\in[0,1]\times[0,d],\\
& u(x,0)=u(x,d)=\rho_0-x^2(1-x)^2,\\
& u(0,y)=u(1,y)=\frac{\rho_1-\rho_0}{d}y+\rho_0,
\end{aligned}
\right.
\end{equation}
where $\rho_0>\rho_1$ are suitable constants, that is numerically solved approximating
first and second derivatives by means of second--order and fourth--order finite difference formulas, respectively \cite{pp}. More in detail, let us consider a discretized domain 
$\Omega_D$, with steps $dx$ and $dy$ along $x$ and $y$ directions, respectively, then, 
$\forall \left(x,y\right) \in \overset{\circ}{\Omega}_D$, it is
\begin{equation}
\begin{aligned}
& \rho_x\left(x,y\right) \approx \frac{\rho\left(x+dx,y\right)-\rho\left(x-dx,y\right)}{2dx}, \\
& \rho_y\left(x,y\right) \approx \frac{\rho\left(x,y+dy\right)-\rho\left(x,y-dy\right)}{2dy}, \\
& \rho_{xx}\left(x,y\right) \approx \frac{\rho\left(x+dx,y\right)-2\rho\left(x,y\right)+\rho\left(x-dx,y\right)}{dx^2}, \\
& \rho_{yy}\left(x,y\right) \approx \frac{\rho\left(x,y+dy\right)-2\rho\left(x,y\right)+\rho\left(x,y-dy\right)}{dy^2}.
\end{aligned}
\end{equation}
The algebraic system, resulting from the evaluation of (\ref{BVP}) in each grid point, is solved by using the 
Matlab\textsuperscript{\texttrademark} routine \texttt{fsolve} with the Levenberg--Marquardt algorithm and tolerance $10^{-4}$; moreover, we take $dx = dy = 0.02$.

In Figure~\ref{fig1}, we present some numerical solutions in the linear case; we show both the plot of the mass density as a function of $x$ and $y$, as well as the corresponding contour plot. In all the plots of Figure~\ref{fig1} the following values for the parameters are used: $\alpha=1.0$, $\beta=-1.2$, $\rho_0=1.4$, $\rho_1=1.3$, $d=3$, $n=-1$. The values of parameters $m$ and $\gamma$ used in each plot are specified in the caption.

On the contrary, in Figure~\ref{fig2}, we present some numerical solutions in the nonlinear case; we show both the plot of the mass density as a function of $x$ and $y$, as well as the corresponding contour plot. In all the plots of Figure~\ref{fig2} the following values for the parameters are used: $\alpha=1.0$, $\beta=-1.2$, $\gamma=-1$, $\rho_0=1.4$, $\rho_1=1.3$, $d=3$. The values of parameters $m$ and $n$ used in each plot are specified in the caption.

We stress that the numerical results above presented are only preliminary, even if the derivation of a single equation for the equilibrium is in our opinion remarkable; we are conscious that the analysis of equilibrium configurations of Korteweg fluids, 
in three space dimensions and using boundary conditions and parameters suggested by experiments,  is worth of being deeply investigated.

\section{Conclusions}
\label{sec:conclusions}

In this paper, we considered the balance equations of a third grade Korteweg--type fluid. After reviewing some recent results \cite{GorRog2021}, where a thermodynamical analysis by means of the extended Liu procedure allowed the authors to derive explicitly the constitutive functions compatible with second law of thermodynamics, we focused on the search of purely mechanical equilibrium configurations at constant temperature. 
Limiting ourselves to a two--dimensional setting, we derived a single scalar partial differential equation whose solutions automatically satisfy the overdetermined system for the mechanical equilibrium of a Korteweg fluid. This condition, that in general is expressed as a highly nonlinear elliptic equation, can be solved by choosing an appropriate boundary condition of Dirichlet type. Some numerical solutions have been obtained both in the simple linear case and in some nonlinear ones. The results here provided are only preliminary but seem to be promising. Future developments are under current investigation in order to obtain  equilibrium configurations suitable to be compared with laboratory experiments. Also, the search of equilibria and their stability in a three--dimensional setting is planned. Finally,  a problem that will be worth of being investigated in the near future is concerned with the existence of stationary solutions with a non-uniform temperature field; in two space dimensions, equations (\ref{equilibriumcond}) give two conditions for two unknowns, say $\rho$ and $\theta$, whereas in three space dimensions we have to solve an overdetermined system of three equations and check their compatibility with the existence of non--trivial solutions.

\section*{Acknowledgments}
Work supported by the ``Gruppo Nazionale per la Fisica Matematica''   (GNFM) of the Istituto Nazionale di Alta Matematica (INdAM), by University of Messina and Kore University of Enna.

\end{document}